\documentclass{ws-procs975x65}

\def\mnras{Mon. Not. Roy. Astron. Soc.}
\def\apjl{ApJL}

\def\prd{Phys. Rev. {\bf D}}
\def\apj{Astrophys. Journal}

\newcommand{\vev}[1]{\langle #1 \rangle}
\newcommand{\vp}{{\bf \hat p}}
\newcommand{\vpp}{{\bf \hat p'}}
\newcommand{\vk}{{\bf \hat k}}

\begin{document}



\title{CMB Anomalies from Relic Anisotropy
\footnote{Talk given by Carlo R. Contaldi at the Eleventh Marcel Grossmann Meeting
on General Relativity, Berlin, July 2006.}}

\author{A. Emir G\"umr\"uk\c{c}\"uo\u{g}lu}

\address{School of Physics and Astronomy, University of Minnesota, Minneapolis, MN 55455, USA
}

\author{Carlo R. Contaldi}

\address{Theoretical Physics, Blackett Laboratory, Imperial College, London, SW7 2BZ, UK
}

\author{Marco Peloso}

\address{School of Physics and Astronomy, University of Minnesota, Minneapolis, MN 55455, USA
}

\begin{abstract}
Most of the analysis of the Cosmic Microwave Background relies on the
assumption of statistical isotropy.  However, given some recent
evidence pointing against isotropy, as for instance the observed
alignment of different multipoles on large scales, it is worth testing
this assumption against the increasing amount of available data.  As a
pivot model, we assume that the spectrum of the primordial
perturbations depends also on their directionality (rather than just on
the magnitude of their momentum, as in the standard case). We
explicitly compute the correlation matrix for the temperature
anisotropies in the simpler case in which there is a residual isotropy
between two spatial directions. As a concrete example, we consider a
different initial expansion rate along one direction, and the
following isotropization which takes place during inflation. Depending
on the amount of inflation, this can lead to  broken statistical
isotropy on the largest observable scales.
\end{abstract}

\bodymatter

\section{Introduction and discussion}\label{intro}
The release of the WMAP 3--year results \cite{spergel06} has now
placed our knowledge of the largest scales of the CMB on a firm
footing. An intriguing aspect of the maps produced by WMAP is the
presence of a number of seemingly related anomalies on the largest
scales. They comprise the lack of power in the lowest multipoles, the
alignment of the power in a number of the lowest multipoles and an
apparent asymmetry in the maps. The lack of power and its possible
origin has been addressed copiously in the literature
\cite{lowell1,contaldi03,lowell2}. The difficulty of calculating the
posterior probability of the measured values of the quadrupole and
octupole has led to a spread in the claims of the significance of the
effect which now appears to be marginal \cite{lowell2}.

The significance of the anomalies which represent a breaking of the
statistical isotropy of the maps appears much stronger
\cite{anomalies,magueijo06}. Indeed,
if the signal {\sl is} statistically anisotropic then much of the
information content of the maps lies in the anisotropic
correlation of the spherical harmonic modes with $\langle a^{\,}_{\ell
  m}a^\star_{\ell' m'}\rangle \propto \!\!\!\!\! \not \;\;\;\;
\delta_{\ell\ell'}\delta_{mm'}$. Obvious explanations for the observed
anomalies are that an unknown
systematic is not being accounted for in the analysis or that
foreground signals, which are naturally anisotropic in nature, are not
being modelled and subtracted properly. However a conclusive
explanation along these lines has not been put forward yet.  

Another, albeit more speculative, possibility is that the explanation
lies in more fundamental aspects of the cosmological model. It seems
advantageous therefore to attempt to model the possible anisotropies
in the sky from first principles from a primordial perspective with a
view to constrain departures from the standard cosmological picture.
Here we assume that the broken statistical isotropy
is imprinted in the primordial spectrum $P_\Phi \left( {\bf k} \right)$
of the cosmological perturbations, leading to a
specific and falsifiable pattern for the covariance
matrix of the CMB
fluctuations \footnote{This is in contrast to the opposite
approach of constraining general modifications to the covariance
\cite{spergel06,magueijo06} such as $\langle a^{\,}_{\ell
  m}a^\star_{\ell' m'}\rangle \propto 
\delta_{\ell\ell'}\delta_{mm'}(C_\ell + \epsilon_{\ell m})$.}.

The present discussion is organized as follows. We first compute the
general correlations that are obtained in this case. We then present a
simple model which gives rise to the above correlations. The
background evolution of this model is axisymmetric, with an expansion
rate which is initially much faster along a single direction. For
simplicity, we assume isotropy in the remaining two directions. A
period of inflation than leads to full isotropy. Such a situation may
emerge in models with extra dimensions (as for instance in string
theory). In this case, one faces the difficult problem to explain why
only three directions undergo cosmological expansion, while the other
remain stabilized on a small size~\cite{brva}. It is plausible to
imagine that the expansion of the three (now) large dimensions did not
start precisely with the same rate, but that isotropy was only reached
afterwards in the first few e-foldings of inflation. With this
approach, matter and radiation perturbations propagate in an
isotropically expanding background. The breaking of the statistical
isotropy is therefore entirely due to the initial conditions of the
model (as opposed to a full anisotropic evolution of perturbations
\cite{bunn96,gordon05}). We conclude by showing the spectrum of
perturbations of scalar modes having the momentum in the anisotropic
direction (as we discuss, the computation is considerably simpler in
this case).

Details of the computations, and the comparison with the observations,
will be presented elsewhere~\cite{progress}.

\section{CMB covariance matrix}

We want to calculate

\begin{equation}
C_{\ell\ell' m m'} \equiv  \vev{a^{\,}_{\ell m}a^\star_{\ell'm'}} = \int
  d\Omega_\vp d\Omega_\vpp \vev{\delta T(\vp,\eta_0,{\bf x}_0)\delta T(\vpp,\eta_0,{\bf x}_0)} Y_{\ell m}^\star(\vp)Y^{\,}_{\ell' m'}(\vpp),
\label{cllmm}
\end{equation}
where $\delta T$ is the temperature perturbation in the
direction $\vp$, as measured by an observer at position ${\bf x}_0$ and at time $\eta_0 \,$; it can be expressed as
\begin{equation}
\delta T(\vp,\eta_0,{\bf x}_0) = \int \frac{d^3{\bf k}}{(2\pi)^3} \Phi({\bf
  k},\eta_i)\Delta(k,\vk\cdot\vp,\eta_0)  e^{i{\bf k}\cdot{\bf x}_0},
\end{equation}

We assume that the anisotropy is only imprinted in the primordial
curvature power spectrum.  Specifically, $(2\pi / k)^3\delta^3(\vec k
-\vec k') P_\Phi(\vec k) \equiv \langle \Phi(\vec k) \Phi^\star(\vec
k)\rangle$ depends also on the orientation of the wavevectors, and not
only on the magnitude.  In contrast, the calculation of the later time
transfer function $\Delta_\ell(k,\eta_0)$ for the radiation
perturbation proceeds as the standard case by the line of sight
integration of the Einstein--Boltzmann system. Note that the transfer
functions are isotropic in that they only depend on the magnitude of
the wavevector $\vec k$. However the final integration over
wavevectors $\vec k$ must be modified due to the directionality
dependence of $P_\Phi(\vec k)$.

For definiteness, we assume a residual symmetry along two
directions. We denote by $\xi$ the cosine of the angle between the wavevector
and the axis of symmetry. Working the expression~(\ref{cllmm}) out, we find
\begin{eqnarray}
C_{\ell \ell' m m'} &=& \frac{\delta_{m m'}}{\pi} \left( - i \right)^{\ell - \ell'} \sqrt{\frac{(2\ell+1)(2\ell'+1)(\ell-m)!(\ell'-m)!}{(\ell+m)!(\ell'+m)!}} \nonumber\\
&& \;\;\;\;\;\;\;\;\; \times \int \frac{d k}{k} \Delta_\ell \left( k , \eta_0 \right) \Delta_\ell \left( k , \eta_0 \right) \int_{-1}^1 d \xi P_\ell^m \left( \xi \right) P_{\ell'}^m \left( \xi \right) P_\Phi \left( k ,\, \xi \right)
\end{eqnarray}
where $\Delta_l$ are the coefficients of the decomposition of the transfer function into Legendre polynomials).

In the isotropic case, $P_\Phi \left( k ,\, \xi \right) \equiv P_\Phi \left( k \right) \,$, we recover the standard result
\begin{equation}
C_{\ell \ell' m m'}^{({\rm iso})} = \frac{2}{\pi} \delta_{\ell \ell'} \delta_{m m'} \int \frac{d k}{k} P_\Phi \left( k \right) \vert \Delta_\ell \left( k ,\, \eta_0 \right) \vert^2
\end{equation}
In general, we obtain a correlation between different multiples. Here, we focus on a axisymmetric model, for which $P_\Phi ( k, - \xi) = P_\Phi ( k, \xi)$.
In this case, the correlation is non-vanishing whenever the difference between $l$ and $l'$ is even (as it follows from the parity properties
of of the associated Legendre polynomials).

\section{Background evolution}

As a specific example,
we consider a homogeneous but anisotropic
background, with an expansion rate along the $x$-direction different
from the other two,
\begin{equation}
ds^2 = -dt^2 + a\left(t\right)^2\,dx^2+ b\left(t\right)^2\,\left(dy^2+dz^2\right)\,.
\label{bck}
\end{equation}
A scalar field in this background leads to the equations
\begin{eqnarray}
&&\dot H+3\,H^2 = V\,,\nonumber\\
&&\ddot \phi+3\, H \, \dot \phi +V^\prime = 0 \,,\nonumber\\
&&3 H^2 - h^2 = \frac{1}{2} \dot{\phi}^2 + V \,,
\label{eqs}
\end{eqnarray}
where $H \equiv \left(H_a +2 \, H_b \right)/3$ and $h \equiv \left(H_a
- H_b \right)/\sqrt{3}$ (we work in units of $M_p = 1$, while prime
denotes differentiation with respect to $\phi$). The first two
equations are identical to the ones obtained in the isotropic case, in
terms of the ``average'' expansion rate $H \,$. The third equation,
appearing as a ``modified'' Friedmann equation, can then be used as an
algebraic equation for the difference $h$ between the two expansion
rates.

\begin{figure}[t]
\begin{center}\epsfig{file=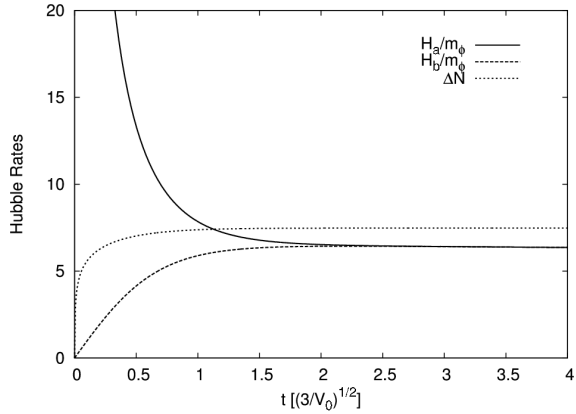,width=7.5cm,angle=0}
\caption{\label{fig:fig1} Evolution of the Hubble rates and the
  difference in e-foldings during the anisotropic and the following
  isotropic inflationary stages. We have chosen a chaotic inflationary
  model, with quadratic potential $V = m_\phi^2 \phi^2 / 2$ with
  initial conditions $\phi_0 = 16 \,, h_0 = 10^4 \, m_\phi \,$. This leads
  to about $60$ e-foldings during the isotropic inflationary stage.} 
\end{center}
\end{figure}

We consider a scalar field potential, and an initial scalar field
value capable of sustaining inflation, and a large initial anisotropy,
$h_0^2 \gg V_0 \gg \dot{\phi_0}^2 \,$. The system then undergoes a
quick anisotropic inflationary stage (with both $\ddot{a} ,\, \ddot{b}
> 0$), during which $\phi$ is practically static, i.e. $V \simeq V_0$,
and the two Hubble parameters behave as $H_a \simeq 1/t$ and $H_b
\simeq 0$.  To follow the evolution of $h \,$, it is useful to combine
eqs.~(\ref{eqs}) into $\dot h + 3 \, H\,h = 0$.
This leads to a very rapid decrease of $h$ during the anisotropic
stage, as the average rate approaches the standard (slow roll) value
$H \simeq \sqrt{V_0 / 3} \,$. The timescale for the decrease is
$t_{\rm iso} \equiv \sqrt{3/V_0} \,$ (we can also see this by
requiring the continuity of $H_a$ between the early $1/t$ behaviour
and the later constant value). From this moment on, $h$ can be
neglected, and we have a second stage of standard (isotropic)
inflation.  Figure~\ref{fig:fig1} shows, for a specific choice of
initial conditions, the evolution of the Hubble parameters along with
the difference in the number of e-foldings between the expansions in
the $x$ and $y$, and $z$ axes, defined as $\Delta N \equiv \int_0 ^t
dt\left(H_a-H_b\right)$. This increases during the first stage and
remains constant after $t_{\rm iso}$.

\section{Perturbations}

The standard FRW background is symmetric under spatial
transformations. It is therefore convenient to classify the
perturbations according to how they transform under spatial change of
coordinates: modes which transform differently are not coupled at the
linearized level. One then finds that $3$ physical modes are
present. They are contained in one scalar perturbation, and in the two
polarizations of a tensor mode, which are also decoupled from each
other.  Also in the present context there are three physical
modes. Two of them are coupled to each other, while the third one is
decoupled, due to the residual symmetry in the $y-z$ plane. The first
two modes decouple when the universe become isotropic, and they become
a scalar and a tensor polarization.

Therefore, we expect that the coupling is proportional to the
difference between the two Hubble rates. We also expect that the
coupling vanishes when the momentum of the modes is along the
privileged direction $x \,$. We denote these modes as
longitudinal. Indeed, the directionality of a mode always breaks the
full $3$d isotropy; when the momentum is oriented along the
anisotropic direction $x$ we have the same symmetries as in the
standard case.  Therefore, the computation is much simpler for these
modes. Here, we focus only
on this simpler case. The complete computation is given
in~\cite{progress}.

We proceed as in the standard case, by writing down the most general
scalar perturbations of the metric and of the inflaton field, and by
choosing a gauge which completely fixes the freedom of coordinates
transformations (see \cite{mfb} for a review). In the longitudinal
case, the linearized Einstein equations for the perturbations can be
reduced to a unique equation in terms of a single variable $Q \,$,
which is the generalization of the Mukhanov-Sasaki
variable~\cite{musa}, and which obeys the equation
\begin{eqnarray}
&& \ddot Q+\left(H_a + 2\,H_b\right)\dot Q+ \left[ \frac{k^2}{a^2} +  {\cal M}^2 \right]\,Q =0 \, \nonumber\\
&& {\cal M} \equiv  V^{\prime\prime}\,+\,2\,\dot\phi\, V^\prime /H_b \,-\,\dot
\phi^4 /\left(2\,H_b^2\right)\,+\,2\,H_a\,\dot \phi^2 /H_b\,+\,\dot
\phi^2
\label{eqQ}
\end{eqnarray}
where $k$ is the comoving momentum, oriented in the $x$
direction. This equation coincides with the standard one when
$H_a=H_b=H$. We note that the scale factor $a$ is entering in the
physical momentum $k/a$, since the mode is moving along the
$x$-direction. The initial conditions are given as in the standard
case, since $k/a$ dominates the last term of eq.~(\ref{eqQ}) at early
times, leading to a standard adiabatic initial vacuum.

\begin{figure}[t]
\begin{center}\epsfig{file=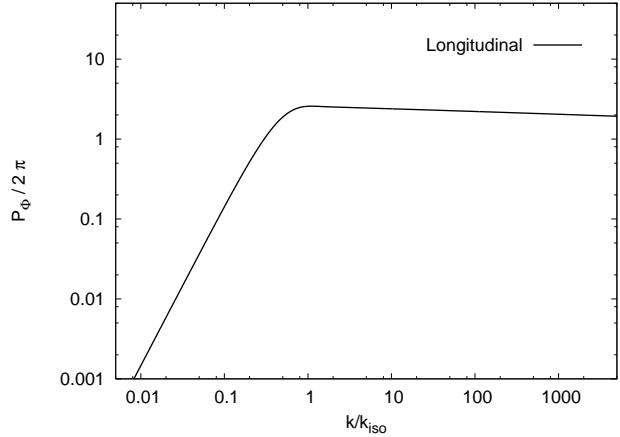,width=6cm,angle=270}
\caption{\label{fig:fig2} Power spectrum for longitudinal scalar perturbations. The initial conditions for the
background are as in figure~\ref{fig:fig1}, while $\phi =3$ at the
moment shown here.}
\end{center}
\label{figspe}
\end{figure}

In figure \ref{figspe}, we show the power spectrum towards the end of
inflation. $k_{\rm iso}$ is the comoving momentum of the modes which
left the horizon when the universe became isotropic. Modes with $k >
k_{\rm iso}$ leave the horizon during the later isotropic stage of
inflation. The standard result is recovered for this modes, with the
typical spectral index of massive chaotic inflation. However, modes
with smaller momenta left the horizon when the universe was still
anisotropic, and a nonstandard result is found in that case.

\section*{Acknowledgements}
We thank Jo\~ao Magueijo for useful
discussions. The work of A.~E.~G. was supported by the Hoff Lu
Fellowship in Physics at the University of Minnesota. The work of
M.~P. was supported in part by the Department of Energy under contract
DE-FG02-94ER40823, and by a grant from the Office of the Dean of the
Graduate School of the University of Minnesota.

\end{document}